\documentclass{aastex631}

\usepackage{color}

\shorttitle{AASTeX v6.3.1 Sample article}
\usepackage{float}
\usepackage{booktabs}
\usepackage{subfigure}
\usepackage{graphicx}
\usepackage{footnote}
\usepackage{multirow}
\usepackage{color}
\usepackage{amsmath}
\usepackage{cases}
\usepackage{appendix}
\graphicspath{{./}{figures/}}

\begin{document}

\title{Strongly Suppressed Cosmic-Ray Diffusion in a Wind-Blown Cavity Environment of Supernova remnant G21.8–3.0}

\author[0009-0003-4873-6770]{Yuan Li}
\affiliation{Tsung-Dao Lee Institute, Shanghai Jiao Tong University, Shanghai 201210, People's Republic of China}

\author[0000-0001-9745-5738]{Gwenael Giacinti}
\affiliation{Université Paris Cité, CNRS, AstroParticule et Cosmologie, Paris, F-75013, France}

\author[0000-0002-0197-470X]{Yang Su}
\affiliation{Purple Mountain Observatory $\&$ Key Laboratory of Radio Astronomy, Chinese Academy of Sciences, Nanjing 210023, PRC}
\affiliation{School of Astronomy and Space Science, University of Science and Technology of China, Hefei, Anhui 230026, PRC}

\author[0000-0002-9392-547X]{Xiao Zhang}
\affiliation{School of Physics $\&$ Technology, Nanjing Normal University, No.1 Wenyuan Road, Nanjing 210023, PRC}

\author[0000-0002-1253-2763]{Hui Li}
\affiliation{Department of Astronomy, Tsinghua University, Beijing 100084, People's Republic of China}

\author[0000-0002-4753-2798]{Yang Chen}
\affiliation{School of Astronomy and Space Science, Nanjing University, Nanjing 210023, PRC}

\email{yuanlss17@sjtu.edu.cn}
\email{ggiacinti@apc.in2p3.fr}
\email{ygchen@nju.edu.cn}

\begin{abstract}

We report the discovery of a GeV $\gamma$-ray source spatially offset from SNR G21.8-3.0, while no significant high-energy emission is detected from the remnant itself. Using 16 yr of \emph{Fermi}-LAT observations together with $^{12}$CO ($J=1\!-\!0$) data, we show that the GeV emission is spatially associated with a localized molecular cloud located just outside a gas cavity whose boundary closely follows the SNR shell. This configuration supports a scenario in which cosmic rays (CRs) escaping from the remnant propagate through a progenitor-carved low-density cavity and illuminate nearby dense gas. Modeling of the cloud emission requires a diffusion coefficient suppressed by a factor of several tens relative to the canonical Galactic value, indicating strongly inhibited CR transport around the remnant. Rather than serving as a simple low-density escape channel, the wind-blown cavity appears to regulate CR propagation through local magnetic structure and enhanced turbulence. This system therefore provides evidence that circumstellar structures can control the escape and early transport of CRs around supernova remnants.

\end{abstract}

\keywords{cosmic rays; supernova remnants; circumstellar matter; interstellar medium;}

\section{Introduction} \label{sec:1}

Massive stars can profoundly reshape their surrounding interstellar medium through stellar winds and ionizing radiation before exploding as supernovae \citep{1975ApJ...200L.107C,1977ApJ...218..377W}. These feedback processes may excavate low-density cavities surrounded by dense gas, thereby modifying both the gas distribution and the subsequent transport of cosmic rays (CRs) released by the supernova remnant (SNR) \citep{1999ApJ...511..798C,2004A&A...424..747P}. In the standard escape scenario, CRs accelerated at SNR shocks diffuse into nearby dense gas and produce $\gamma$-ray emission displaced from the remnant itself \citep{2009MNRAS.396.1629G}. As interactions between SNRs and molecular clouds are widely observed in the Galaxy \citep{2010ApJ...712.1147J}, well-known examples such as W28 and W44 have shown that spatially offset $\gamma$-ray emission can trace nearby molecular material illuminated by escaped CRs \citep{aharonian2008discovery,uchiyama2012fermi}. Some systems show bright $\gamma$-ray emission from both the SNR and nearby molecular clouds, with the external emission interpreted as arising from escaped CRs \citep{2023ApJ...953..100L,2023ApJ...945...21L,2024A&A...689A.257L}. However, systems in which the SNR shell is faint in high-energy $\gamma$-rays while only localized external dense gas is bright remain comparatively rare \citep{2015ApJ...809..102L}. Such sources are particularly valuable because they probe not only CR escape from the shock, but also how the circumstellar environment regulates the early propagation of CRs. 

The SNR G21.8–3.0 (G21.8 hereafter) was recently identified as an extended shell‑like remnant in radio continuum surveys, exhibiting a non‑thermal spectrum with a spectral index $\alpha\approx-0.72$ and significant polarized emission characteristic of synchrotron radiation from shock‑accelerated electrons\citep{2020MNRAS.493.2188G}. Radio observations from Urumqi$\lambda-$6cm, Effelsberg$\lambda-$11cm, and Effelsberg$\lambda-$21cm revealed a roughly circular morphology spanning about one degree, with elongated filamentary H$\alpha$ features along the northern edge \citep{2001PASP..113.1326G}. To date, no X-ray counterpart has been detected in the \emph{ROSAT} and \emph{Chandra} data archives \citep{1999A&A...349..389V,2020MNRAS.493.2188G}, nor has any TeV emission been detected within the SNR region by HAWC and HESS \citep{2019ICRC...36..706J}.

In this work, we report such a configuration toward SNR G21.8 using 16 yr of \emph{Fermi}-LAT data together with $^{12}$CO ($\rm J=1\!-\!0$) observations from the Milky Way Imaging Scroll Painting (MWISP\footnote{\url{http://english.dlh.pmo.cas.cn/ic/}}). We detect a GeV $\gamma$-ray source adjacent to the remnant, whereas no significant $\gamma$-ray emission is found from the SNR shell itself. The GeV emission is spatially associated with a nearby molecular cloud just outside a gas cavity whose boundary closely matches the SNR shell, suggesting that the remnant is evolving in a progenitor-carved low-density environment. This morphology supports a scenario in which CRs escaping from the SNR propagate through the cavity and illuminate the surrounding dense gas. By modeling the cloud emission, we find that the diffusion coefficient around the remnant is significantly suppressed relative to the Galactic average, implying that the cavity acts not as a transparent channel for CR escape, but as a turbulent environment that regulates CR transport. Our results suggest that circumstellar structures shaped by massive-star feedback can play an important role in controlling the escape and early propagation of CRs around SNRs.

\section{Observations}\label{sec:2}
\subsection{GeV observation results}
In this part, Fermi-LAT Pass 8 data events (evclass=128 and evtype=3) from August 4, 2008 (Mission Elapsed Time 239557418) to August 4, 2024 (Mission Elapsed Time 744465605) are adopted. The events was collected from 0.1 - 1000 GeV energy range within a 14$^\circ$ $\times$ 14$^\circ$ region of interest (ROI) centered on 4FGL J1840.4-1139 \citep{Abdollahi2020a,2023arXiv230712546B}. The events with zenith angles $>$ 90$^\circ$ were excluded to avoid the contamination from Earth limb, the standard data quality criteria $\tt (DATA\_QUAL > 0)  \&\& (LAT\_CONFIG == 1)$ also considered. The standard LAT analysis package Fermitools v2.2.0 and Fermipy v1.1.6 \citep{2017ICRC...35..824W} are adopted in the subsequent analysis. The binned maximum likelihood analysis \citep{mattox1996likelihood} is adopted together with two diffuse background models (IEM, $\tt gll\_iem\_v07.fits$)/($\tt iso\_P8R3\_SOURCE\_V3\_v1.txt$ ). The likelihood test statistic (TS) is adopted to calculated the significance of the $\gamma$-ray sources, which is defined as TS$= 2 (\ln\mathcal{L}_{1}-\ln\mathcal{L}_{0})$, where $\mathcal{L}_{1}$ and $\mathcal{L}_{0}$ are maximum likelihood values for the model with and without target source. Furthermore, the $\rm TS_{ext}$ is defined as $\rm {TS_{ext}} = 2(\ln\mathcal{L}_{\rm ext} - \ln\mathcal{L}_{\rm ps})$, where $\mathcal{L}_{\rm ext}$ and $\mathcal{L}_{\rm ps}$ represent the maximum likelihood values for the extended and point-like templates, respectively.

\begin{figure*}
    \centering
    \includegraphics[trim={0 0.cm 0 0}, clip,width=0.325\textwidth]{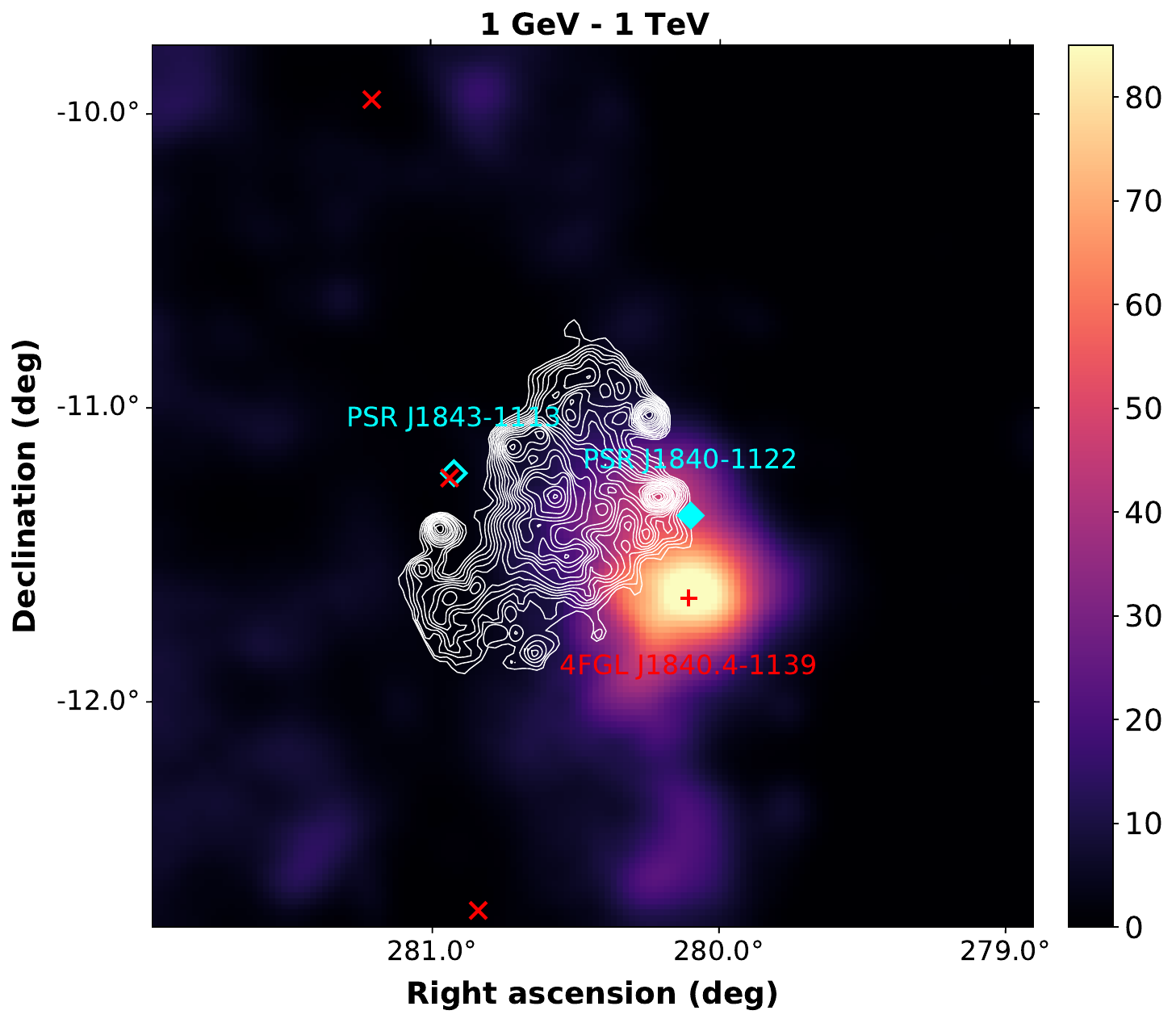} 
    \includegraphics[trim={0 0.cm 0 0}, clip,width=0.325\textwidth]{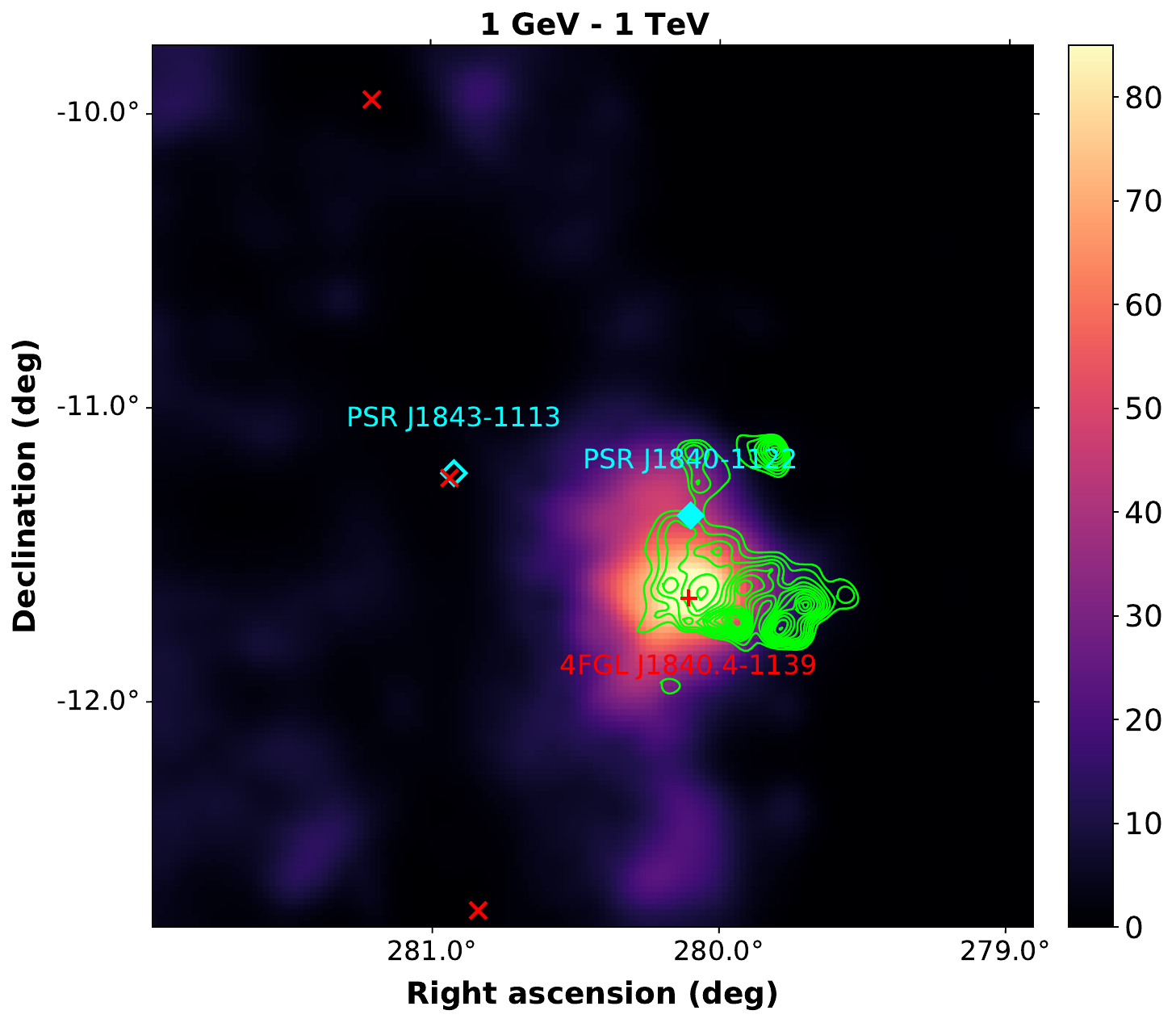} 
    \includegraphics[trim={0 0.cm 0 0}, clip,width=0.325\textwidth]{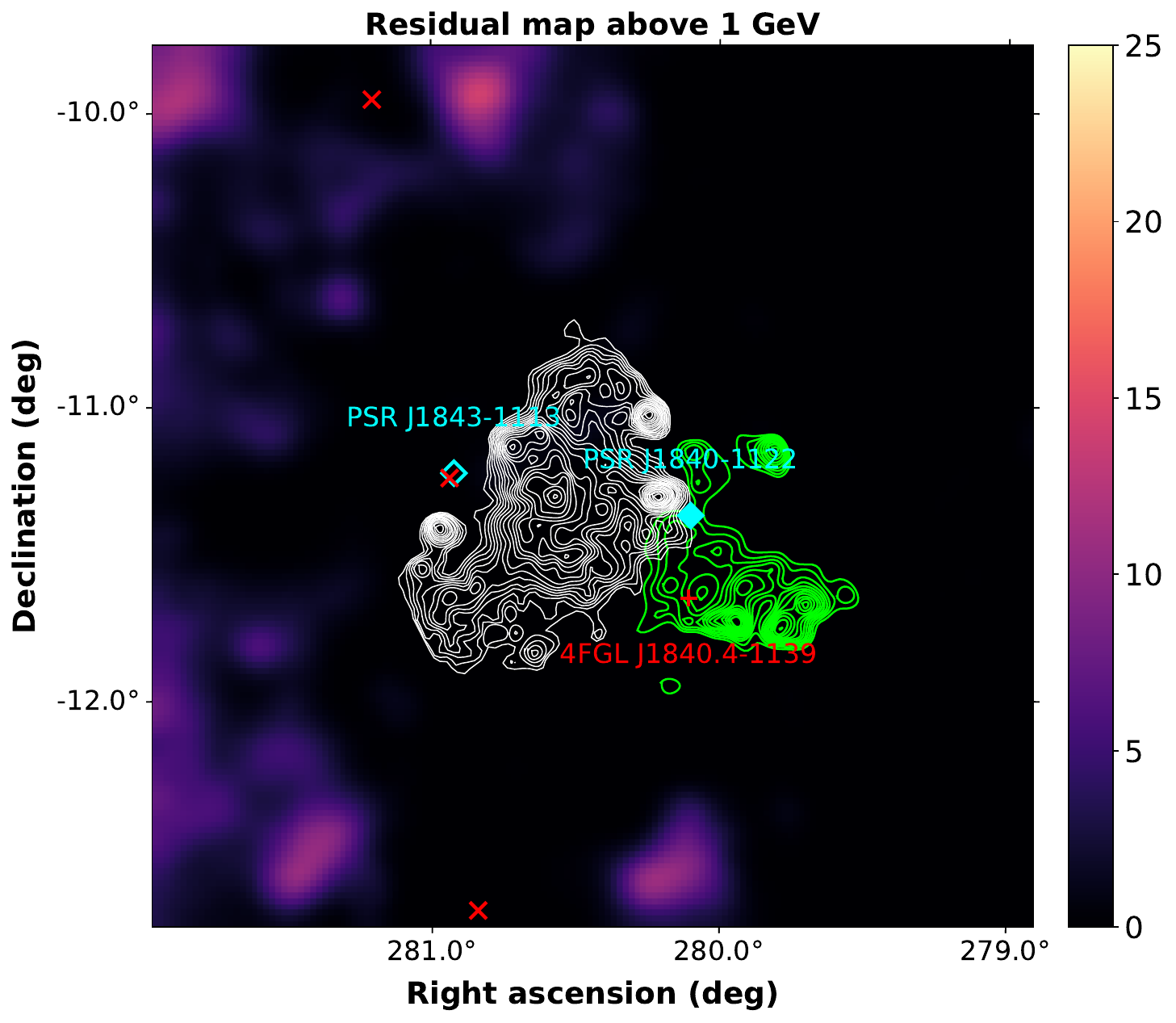} \\
    \includegraphics[trim={0 0.cm 0 0}, clip,width=0.45\textwidth]{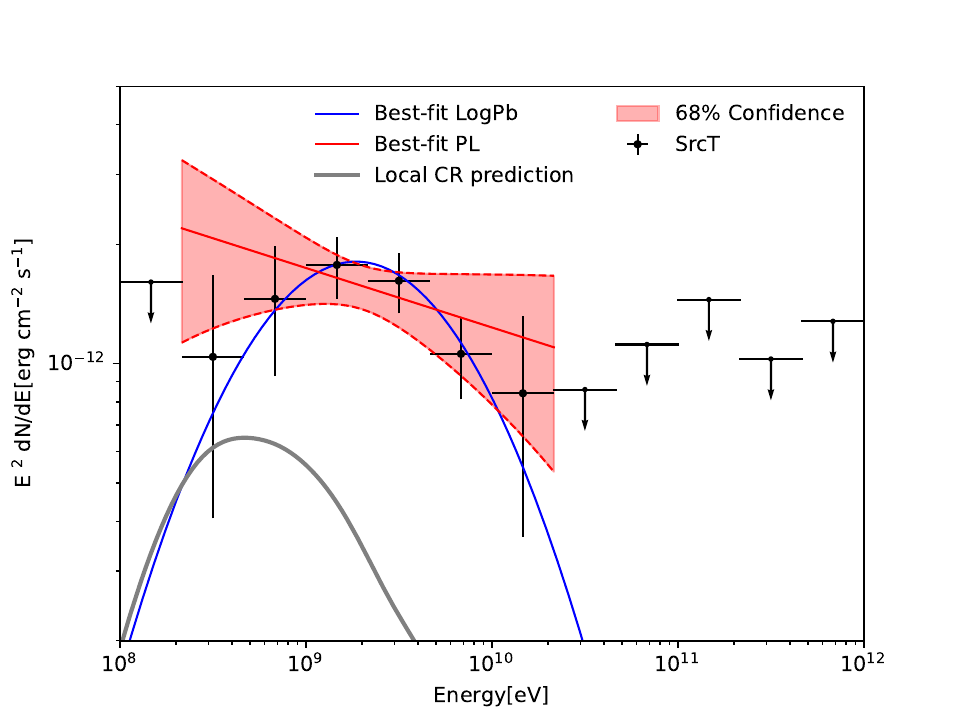} 
    \includegraphics[trim={0 0.cm 0 0}, clip,width=0.5\textwidth]{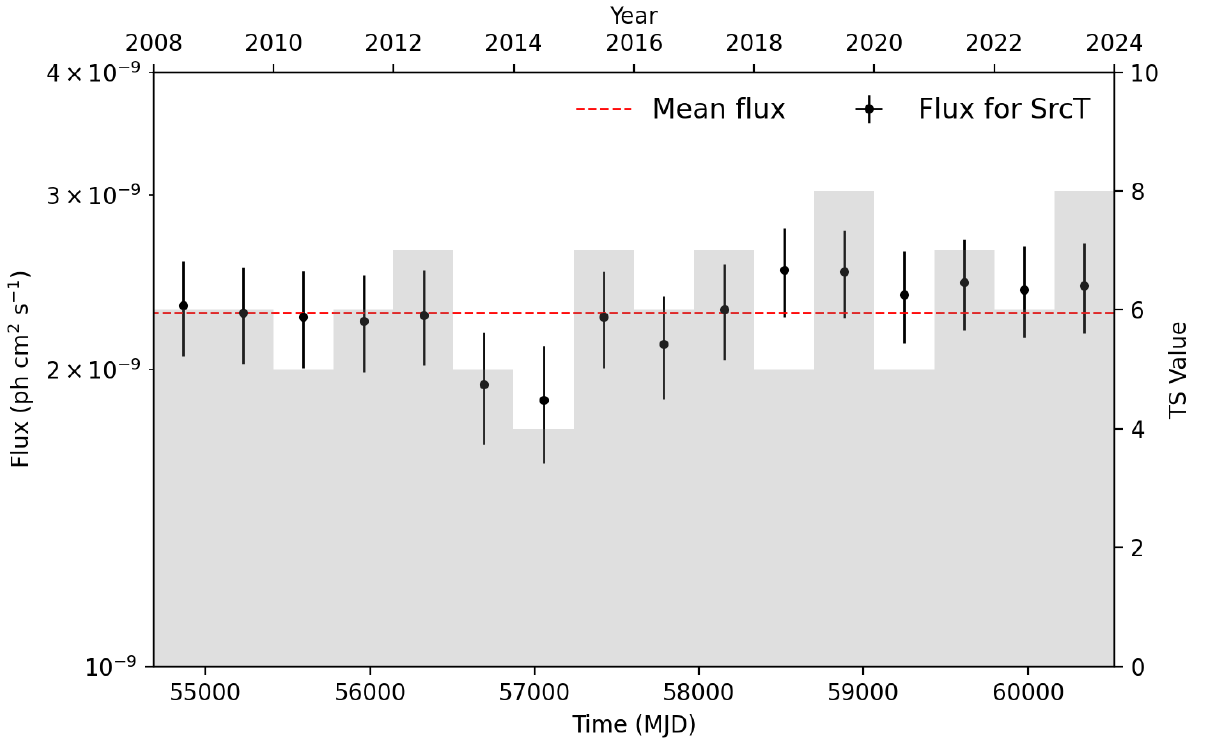} \\ 
 
    \caption{\textbf{Top row}: 3$^{\circ}\!$.0 $\times$3$^{\circ}\!$.0 TS map around SNR G21.8 region in the energy range of 1 GeV - 1 TeV. In the left and middle panels, all diffuse background and 4FGL-DR4 background sources except for 4FGL J1840.4-1139 (SrcT) are subtracted. The location of SrcT is shown as the red plus, and the background sources in 4FGL-DR4 are shown as red crosses. Two energetic pulsar's position shown as the cyan diamond \citep{2005AJ....129.1993M}. The white contour show the radio emission of SNR G21.8 from the Effelsberg $\lambda$ 11-cm observations \citep{2020MNRAS.493.2188G}. The green contours show the $^{12}$CO (J= 1-0) emission obtained from the MWISP measurement (details can be seen in Section \ref{sec:gas}. \textbf{Bottom left}: The SED of SrcT, the black data points are calculated by Fermi-LAT in the energy range of 100 MeV - 1 TeV. The black arrows indicate the 95$\%$ upper limits. The red dashed lines and butterfly show the best-fit PL with 1$\sigma$ statistical errors. The blue solid line represents the best-fit LogPb spectrum. The gray solid line represents the predicted $\gamma$-ray emission, assuming that the CR spectra therein is the same as the one measured locally by AMS-02 \citep{2015PhRvL.115u1101A}. \textbf{Bottom right}: Light curve of SrcT from 0.1 GeV - 10 GeV, the mean flux and TS value for each time bin are shown as the red dashed horizontal line and gray histogram, respectively.}
   \label{fig:1}
 \end{figure*}

After removing the 4FGL J1840.4-1139 from the background model, we generated TS maps above 1 GeV, as shown in the left and middle panels of Figure~\ref{fig:1}. An additional point-like source was then introduced into the model, and its best-fit position (hereafter SrcT) was determined using the \texttt{gtfindsrc} tool. The derived location is consistent with that reported in the 4FGL catalog, the 68$\%$ error radii $\rm {R_{68}}$ is around $\theta_{68}=0.08^\circ$. To test for possible spatial extension, we replaced SrcT with both a uniform disk and a two-dimensional Gaussian template using \texttt{Fermipy}. The resulting extension test statistics are low, with $\rm TS_{ext} \sim 2$ and 3, respectively, indicating that no significant extension is detected. We further performed spectral modeling in the 0.1-1000 GeV energy range using both a power-law (PL) and a log-parabola (LogPb) model. The LogPb model provides a significantly better fit than the PL model, and the $\rm {TS_{curve}}$ value defined as $\rm {TS_{curve}} = 2(\ln\mathcal{L}_{\rm LogPb} - \ln\mathcal{L}_{\rm PL}) = 46.7$, which corresponds to a significance level of 6.8$\sigma$ with one additional degree of freedom \citep{Abdollahi2020a}, suggesting the presence of spectral curvature. The spectrum of SrcT is well described by a LogPb with alpha =0.59 $\pm$ 0.18 and a significantly curvature of beta = 0.28$\pm$0.06, the integrated photon flux from 100 MeV to 1 TeV is calculated to be (1.35$\pm$0.27)$\times$10$^{-9}$ photon cm$^{-2}$ s$^{-1}$. The derived spectral energy distribution are shown in the bottom left of Figure \ref{fig:1}.

To test the possibility that SrcT is a point-like variable $\gamma$-ray source (e.g. AGN), we tested the flux variability. The photon events with energies between 0.1 - 10 GeV were divide into sixteen time bins. For the timing analysis, all spectral indices of the sources were fixed to the best-fit values, only normalizations are allowed to vary, then we repeated the likelihood analysis in each time bin. The derived light curve are shown in the bottom right of Fig.\ref{fig:1}. We also calculated the variability index as defined in \citet{2012ApJS..199...31N}, which under the assumption that with a value in the null hypothesis where the source flux is constant across the full time period, and the value under the alternate hypothesis where the flux in each bin is optimized as TS$_{\rm var}$ = $\sum_{i=1}^{N} 2 \times (\rm{Log}(\mathcal{L}_i(\mathcal{F}_i))-\rm{Log}(\mathcal{L}_i(\mathcal{F}_{\rm mean})))$, where $\mathcal{L}_i$ is the likelihood corresponding to bin $i$, $F_i$ is the best-fit flux for bin $i$, and $\mathcal F_{\rm mean}$ is the best-fit flux for the full time assuming a constant flux. For this LC, the source of TS$_{\rm var} \geq 16.81$ was considered to be a variable source with a 99$\%$ confidence level, given the TS$_{\rm var}$ is anticipated to be distributed as $\chi^2_{N-1}(\rm TS_{var})$ in the null case, we get TS$_{\rm var}$ = 7.9 for the flux variability using above energy band, which is very marginal, thus we suggests no significant variability for the $\gamma$-ray excess from SrcT. Nevertheless, we notice that the TS values in the seventh time bins show a slight decline. Since the data points lie within the 1 $\sigma$ uncertainty region, the photon flux can be considered stable. The evidence is therefore insufficient to draw a firm conclusion about flux variability, which can be attributed to statistical fluctuations.

\subsection{CO observation results}\label{sec:gas}

\begin{figure*}
    \centering
    \includegraphics[trim={0 0.cm 0 0}, clip,width=0.4\textwidth]{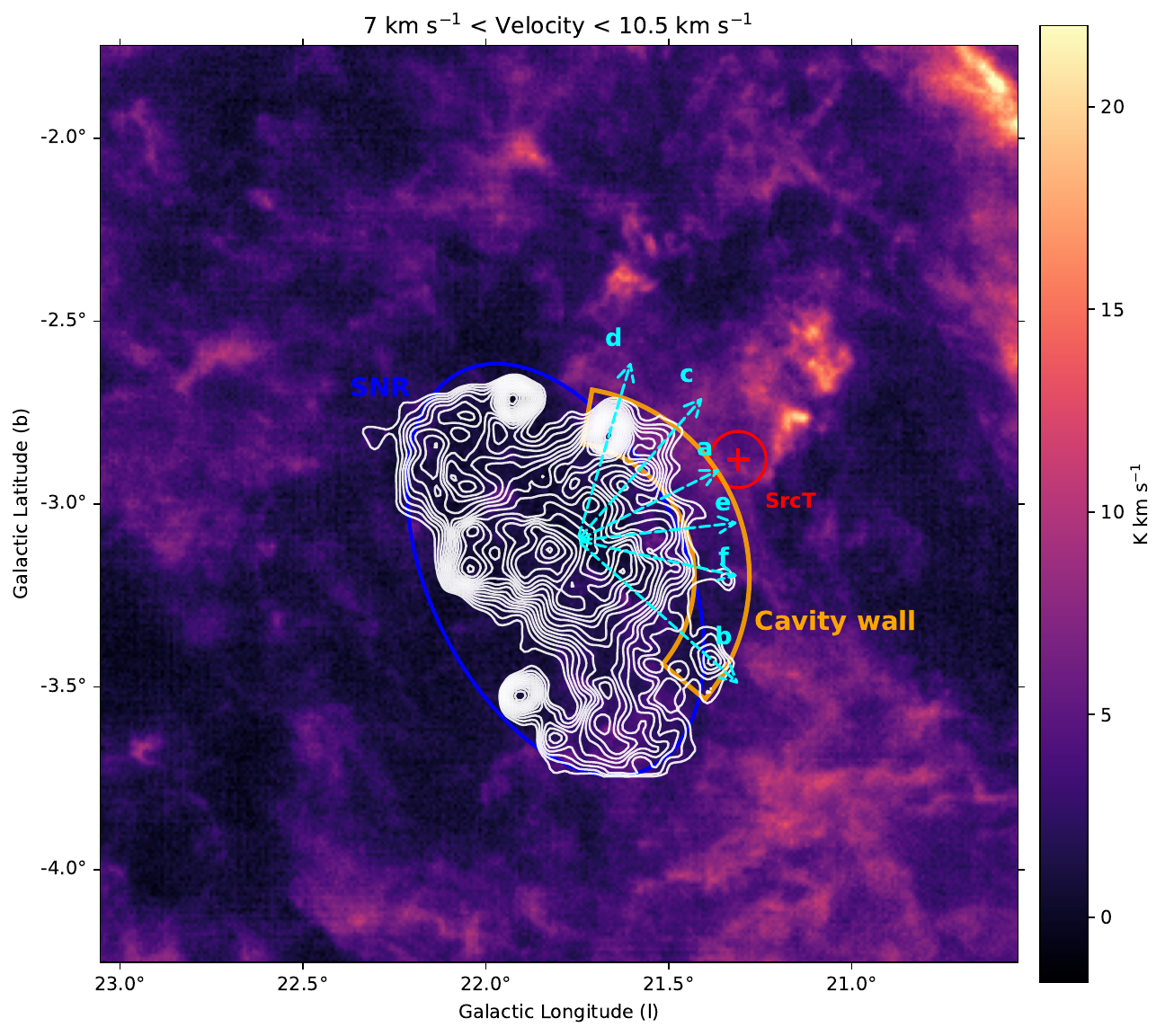} 
    \includegraphics[trim={0 0.cm 0 0}, clip,width=0.45\textwidth]{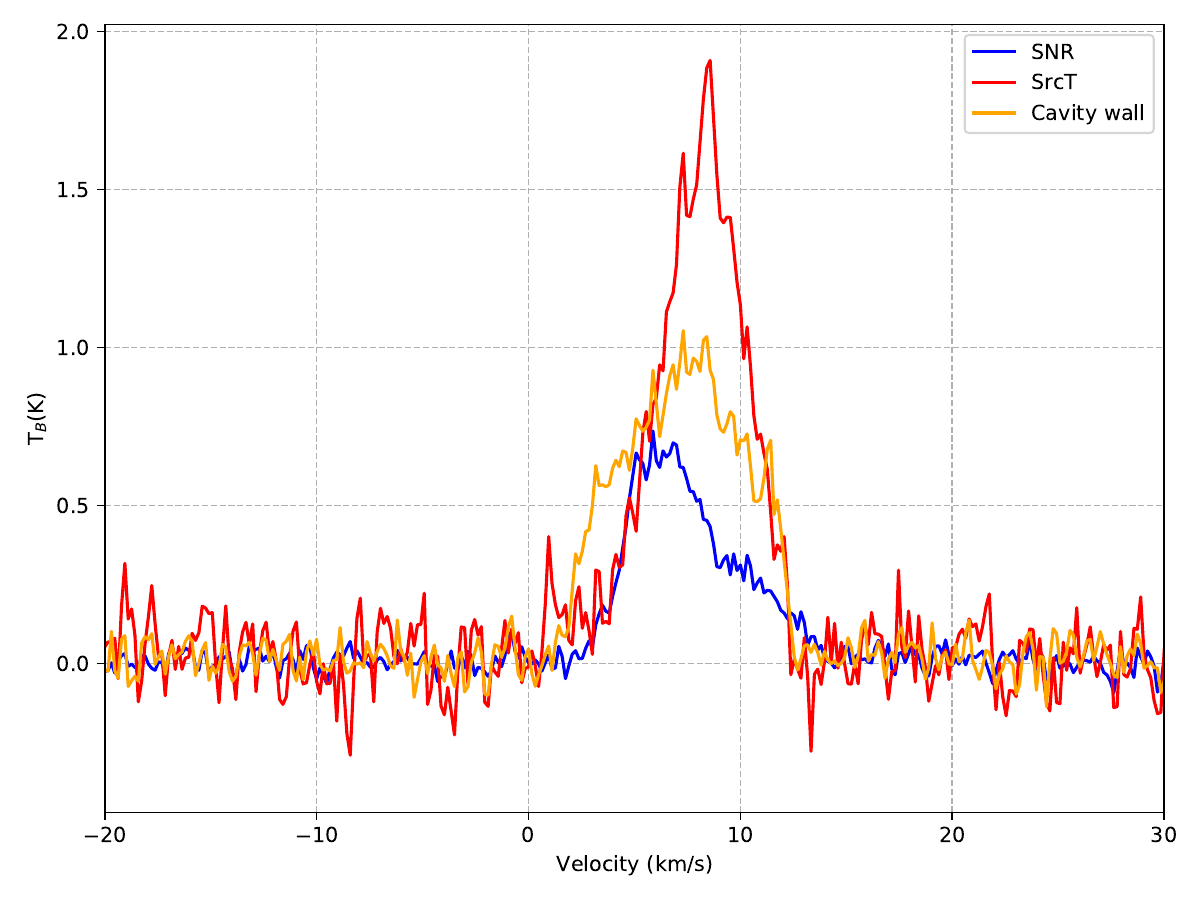}\\
    \caption{Left panel: integrated $^{12}$CO (J= 1$-$0) emission intensity (K km $\rm s^{-1}$) in the velocity range of 7 $-$ 10.5 km $\rm s^{-1}$. The white contours and the red cross are the same as in Figure \ref{fig:1}. The blue ellipse represents the roughly position and extension of SNR. The different cyan annotations show the different directions to extract position-velocity diagrams, details can be seen in Figure \ref{fig:4}. Right panel:$^{12}$CO (J= 1$-$0) spectra inside each region noted in left panel.}
   \label{fig:3}
 \end{figure*}

\begin{figure*}
    \centering
    \includegraphics[trim={0 0.cm 0 0}, clip,width=0.9\textwidth]{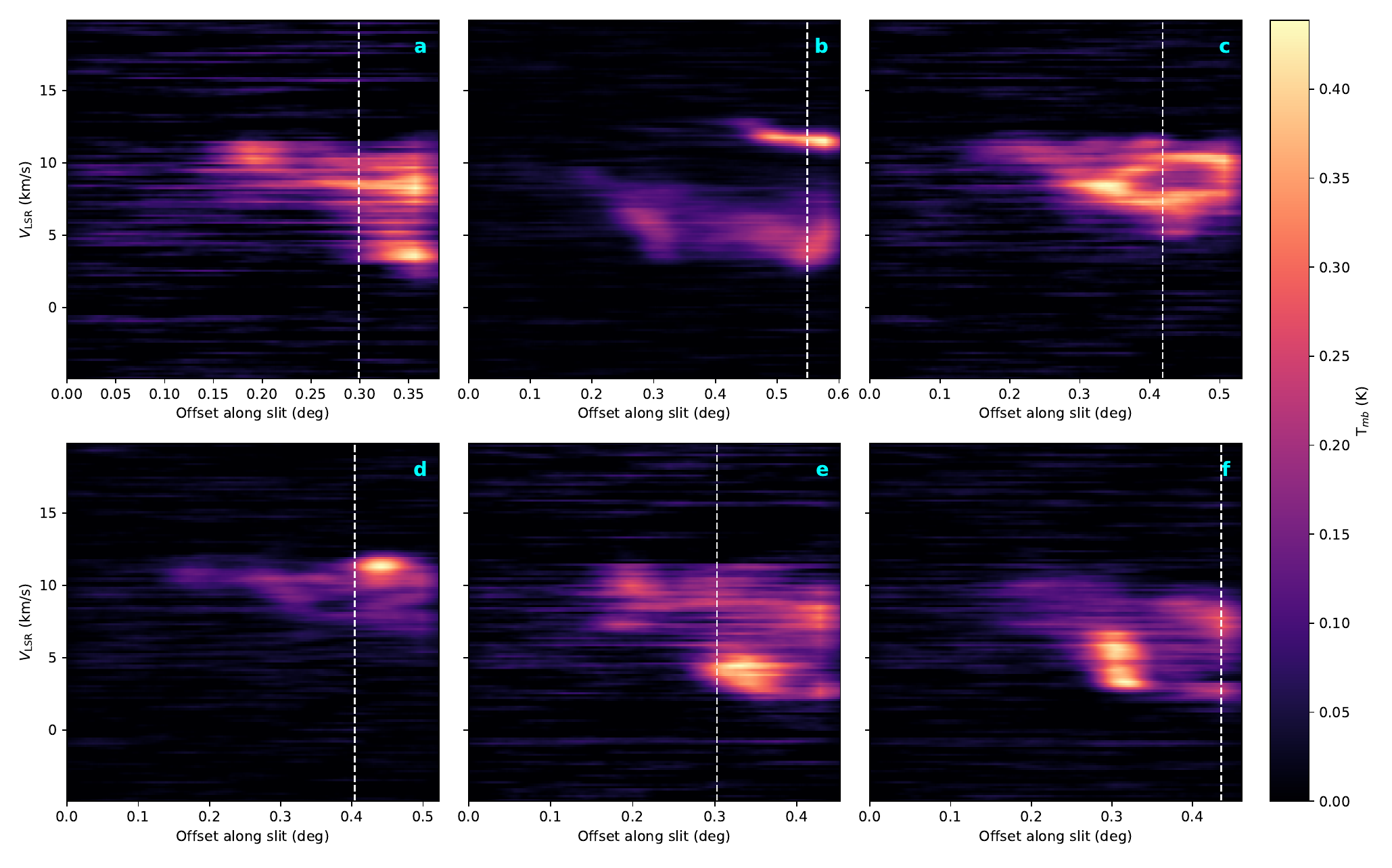}
    \caption{Position-velocity (PV) diagrams of $^{12}$CO (J= 1-0) emission along different directions from $a$ to $f$ in Figure \ref{fig:3}. The white dashed lines corresponds to the SNR radio shell boundary.}
   \label{fig:4}
 \end{figure*}

To trace the molecular gas toward SNR G21.8, we make use of the MWISP CO survey of the northern Galactic plane obtained with the Purple Mountain Observatory (PMO) 13.7 m telescope \citep{2019ApJS..240....9S}. The integrated $^{12}$CO emission around \(v_{\rm LSR}\sim + 9\ \mathrm{km\ s^{-1}}\) reveals a molecular structure surrounding the remnant, with a low-intensity cavity approximately enclosed by brighter molecular gas. A localized molecular cloud is also found close to the GeV $\gamma$-ray excess (Shown in the left panel of Figure \ref{fig:3}). This morphology suggests that the SNR may be located inside a gas cavity and that the $\gamma$-ray source may be associated with dense gas near the cavity boundary. To further test whether the molecular cavity around \(v_{\rm LSR}\sim +9\ \mathrm{km\ s^{-1}}\) is physically associated with SNR G21.8, we analyzed position-velocity (PV) diagrams along several cuts crossing the remnant, labeled from $a$ to $f$ in Figure~\ref{fig:3} and Figure~\ref{fig:4}. These PV diagrams allow us to examine the kinematic structure of the molecular gas relative to the projected SNR shell. Among the directions tested, the cuts along directions $a,b,e$ and $f$ show tentative velocity-broadened features with widths of \(\sim 8\ \mathrm{km\ s^{-1}}\). If interpreted as the expansion of a swept-up molecular shell, this corresponds to a characteristic velocity of a few \(\mathrm{km\ s^{-1}}\). Such broadening may partially reflect the interaction between the SNR or its progenitor's wind and the surrounding molecular material, although turbulence, projection effects, or internal cloud motions may also contribute. Therefore, the PV features provide tentative kinematic support for an association between the molecular cavity and SNR G21.8 or its progenitor.

We then estimate the gas properties of the molecular cloud and cavity using the same CO velocity interval. To determine the H$_2$ column density, we adopt the CO-to-H$_2$ conversion factor \(X_\mathrm{CO}=2\times10^{20}\ \mathrm{cm^{-2}\ K^{-1}\ km^{-1}\ s}\) from \citet{Bolatto2013}. The column density is calculated as $N_\mathrm{H_2}=X_\mathrm{CO} W_\mathrm{CO}$. The molecular gas mass is then derived from $M = \mu m_{\rm H} d^2 \Delta\Omega_{\rm px} X_{\rm CO} \sum_{\rm px} W_{\rm CO}$, where \(\mu=2.8\) accounts for the contribution of helium, \(m_{\rm H}\) is the hydrogen mass, \(d\) is the source distance, and \(\Delta\Omega_{\rm px}\) is the solid angle subtended by each pixel. $W_{\rm CO}=\int T_{\rm mb}(v)\,dv$ is the velocity-integrated CO brightness temperature in each pixel over the adopted velocity interval, where $T_{\rm mb}$ is the main-beam brightness temperature, and $\sum_{\rm px}$ denotes summation over all pixels within the selected sky region. Since the GeV source is unresolved by \emph{Fermi}-LAT, we estimate the gas properties within the \(\rm R_{68}\) containment region of the source, corresponding to \(\theta_{68}=0.08^\circ\). Then we adopt a fiducial distance of $d=0.8$~kpc, close to the near kinematic distance solution for the molecular gas at $v_{\rm LSR}\sim+9~\mathrm{km\,s^{-1}}$ \citep{2014ApJ...783..130R}, and define $d_{0.8}\equiv d/(0.8\ \mathrm{kpc})$. This estimate is subject to substantial uncertainties due to non-circular motions, and the association between the molecular gas and the SNR is assumed. Assuming a source distance of \(d=0.8\ \mathrm{kpc}\), the total molecular gas mass within the SrcT region is estimated to be M $\sim 101\,d_{0.8}^{2}\ M_\odot$. Approximating the emitting region as a sphere with radius \(R=d \times \theta_{68}\), we derive an average molecular hydrogen density of
$n \sim 253\,d_{0.8}^{-1}\ \mathrm{cm^{-3}}$. This spherical approximation is used only to obtain an order-of-magnitude average density, since the actual molecular gas distribution is irregular. On larger scales, the SNR appears to reside within a low-density molecular cavity enclosed by denser gas. The mean molecular hydrogen density in the cavity region is estimated to be $n_{\rm cavity}\sim 8\,d_{0.8}^{-1}\ \mathrm{cm^{-3}}$. The cavity wall is defined by an annular sector centered at $(l,b)=(21.8^\circ,-3.2^\circ)$, with inner and outer angular radii of $0.37^\circ$ and $0.52^\circ$, respectively, and an azimuthal range of $320^\circ$--$80^\circ$ (an opening angle of $120^\circ$). Assuming a line-of-sight depth equal to its projected radial thickness, $L=R_{\rm out}-R_{\rm in}$, we estimate the mean molecular hydrogen density as $n_{\rm wall}=M_{\rm wall}/(\mu m_{\rm H} A L)\sim 82\,d_{0.8}^{-1}\ \mathrm{cm^{-3}}$, where $A$ is the projected area of the sector. The density contrast between the low-density cavity and the localized dense cloud near the GeV excess suggests that the observed $\gamma$-ray emission is likely dominated by interactions between escaping CRs and dense molecular gas outside the cavity.

\section{The origins of gamma rays}\label{sec:3}
\subsection{Progenitor and its wind-driven bubble}\label{sec:bubble}

A search for massive-star counterparts in the vicinity of the $\gamma$-ray excess using the SIMBAD database reveals only a few OB-type stars, including LS~IV~$-$11~36, LS~IV~$-$11~37, and the Be star HD~172252 \citep{2021A&A...649A...1G}, located close to the position of SrcT. Although these objects indicate the presence of massive stars in the region, their small number does not support the existence of a rich massive stellar cluster. Previous studies have shown that $\gamma$-ray emission associated with massive stellar clusters can be powered by the collective effects of many OB stars \citep{2019NatAs...3..561A}, as proposed for Westerlund~1 and Westerlund~2 \citep{2012A&A...537A.114A,2018A&A...611A..77Y}, with further candidate associations reported in the G213 and G172 regions \citep{2025ApJ...990....9L,2026ApJ..1003..100L}. Therefore, the currently identified stellar population appears insufficient to sustain a cluster-driven scenario, favoring instead an origin related to a single, or a few, massive progenitor stars. We note, however, that this conclusion is limited by the incompleteness of current catalogs and by the possible presence of obscured massive stars.

The molecular gas around $v_{\rm LSR}\sim+9\ \mathrm{km\ s^{-1}}$ exhibits partially elliptical structures in the PV cuts along directions $a$ and $e$ of Figure~\ref{fig:4}, close to the boundary of the SNR shell marked by the white dashed lines. The velocity-broadened features deviate from the systemic velocity by approximately \(4.5\)--\(5\ \mathrm{km\ s^{-1}}\), which we adopt as a characteristic expansion velocity of the molecular shell. Such a kinematic feature suggests that the molecular gas may form a swept-up shell surrounding the cavity. This interpretation is consistent with the SNR shock expanding inside a pre-existing wind-blown bubble and approaching, or beginning to interact with, the surrounding molecular material, as also proposed for other SNRs \citep{2025MNRAS.537..464S,2025ApJ...990..213H}. This picture is also consistent with the absence of detectable X-ray emission from the cavity region. The low gas density inferred in Section~\ref{sec:gas} would reduce the emission measure of shocked plasma and hence suppress the thermal X-ray surface brightness. We caution, however, that the lack of X-ray emission may also be affected by absorption and observational sensitivity. Under the assumption that the observed molecular shell traces a wind-blown bubble created by the progenitor, we can estimate the characteristic timescale and wind power using the self-similar solution of \citet{1977ApJ...218..377W}. In this solution, the cavity radius scales as $R_{\rm c} \propto \left(\frac{L_{\rm w}}{\rho_0}\right)^{1/5} t_{\rm c}^{3/5}$, where \(L_{\rm w}\), \(\rho_0\), and \(t_{\rm c}\) are the wind kinetic luminosity, ambient mass density, and dynamical timescale of the bubble, respectively. Since \(v_{\rm c}=dR_{\rm c}/dt=(3/5)R_{\rm c}/t_{\rm c}\), the dynamical timescale can be estimated as $t_{\rm c}=\frac{3}{5}\frac{R_{\rm c}}{v_{\rm c}}
\simeq 9\times10^{5}
\left(\frac{R_{\rm c}}{7.0\,\mathrm{pc}}\right)
\left(\frac{v_{\rm c}}{4.5\,\mathrm{km\,s^{-1}}}\right)^{-1}
\mathrm{yr}$. Here this timescale should be interpreted as the dynamical age of the wind-blown shell, rather than the current age of the SNR. Following the same wind-bubble scaling, the required wind luminosity can be written as
$L_{\rm w}\sim 8\times10^{34}
\left(\frac{R_{\rm c}}{7.0\,\mathrm{pc}}\right)^2
\left(\frac{v_{\rm c}}{4.5\,\mathrm{km\,s^{-1}}}\right)^3
\left(\frac{n_{\rm wall}}{82\,\mathrm{cm^{-3}}}\right)
\mathrm{erg\,s^{-1}}$,
where \(n_{\rm wall}\) is used as a characteristic density of the swept-up molecular material. Therefore, this estimate should be regarded as an order-of-magnitude value. The corresponding mass-loss rate of the progenitor is inferred from $L_{\rm w}=\frac{1}{2}\dot{M}v_{\rm w}^{2}$, where \(v_{\rm w}\) is the stellar-wind velocity. For a characteristic main-sequence OB-star wind speed of \(v_{\rm w}\sim 1000\)--\(2000\ \mathrm{km\ s^{-1}}\), we obtain $\dot{M}\sim (0.6-3)\times10^{-7}\ M_\odot\ \mathrm{yr^{-1}}$. This mass-loss rate is compatible with winds from late-O/early-B type massive stars, broadly corresponding to progenitor initial masses of order \(15\)--\(30\,\rm M_\odot\) \citep{2001A&A...369..574V,2013ApJ...769L..16C}. These estimates support a massive-star core-collapse origin for SNR G21.8 and favor the interpretation that the observed molecular cavity was shaped by the progenitor wind.

\subsection{Escaped Cosmic Rays and the Origin of the $\gamma$-ray Excess}

\label{sec:escape}

\begin{figure*}[htbp]
    \centering
    \includegraphics[trim={0 0.cm 0 0}, clip,width=0.5\textwidth]{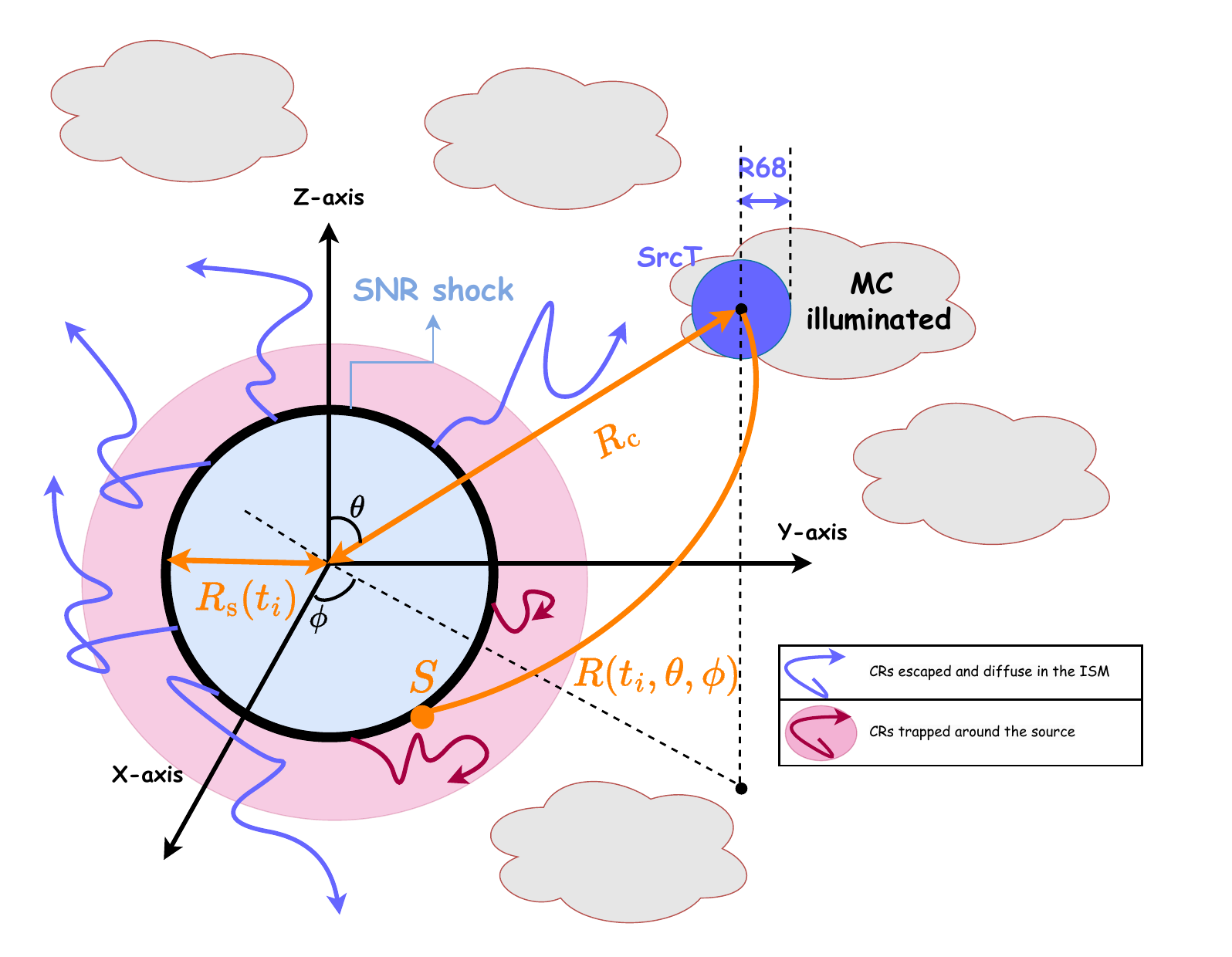}
    \caption{
    Schematic geometry of the escaping-CR scenario. At the escape time $t_i$, particles are released from a source point $S$ on the spherical shock surface of radius $R_s(t_i)$. The centroid of the illuminated $\gamma$-ray emitting region is located at a distance $R_c$ from the SNR center. For an arbitrary field point within the illuminated layer, the distance between the source point $S$ and the field point is written as $\mathcal{R}(t_i,\theta,\phi)$. The characteristic diffusion length after escape is $R_{\rm dif}=2\sqrt{D(E_p)(t-t_i)}$.
    } 
    \label{fig:5}
\end{figure*}

In this section, we consider a scenario in which relativistic protons escape continuously from a spherical shock front during the SNR evolution and subsequently illuminate nearby molecular gas. Although the radio morphology suggests that the remnant is not perfectly spherical, we adopt a spherically symmetric shock as an effective first-order approximation for describing the average distribution of escaped cosmic rays and the resulting $\gamma$-ray emission \citep{1996A&A...309..917A}. For a source point $S$ on the shock surface at the escape time $t_i$, the shock radius is denoted by $R_s(t_i)$. The centroid of the illuminated gas layer is assumed to be located at a distance $R_c$ from the SNR center, which is approximated here by the distance from the SNR center to the centroid of the observed $\gamma$-ray excess. For an arbitrary field point inside the illuminated layer, its radial distance from the SNR center is denoted by $R$. The source-to-field distance can then be written as \citep{2010MNRAS.409L..35L,2012MNRAS.421..935L}:
\begin{equation}
\mathcal{R}(t_i,\theta,\phi)=
\left[
R_s^2(t_i)+R^2-2R_s(t_i)R\sin\theta\cos\phi
\right]^{1/2},
\end{equation}
where $\theta$ and $\phi$ specify the angular position of the source point on the spherical shock surface. The contribution from an elemental patch on the shock surface to the field point follows the standard diffusion Green's function, with the characteristic diffusion length $R_{\rm dif}=2\sqrt{D(E_p)(t-t_i)}$, where $D(E_p)$ is the energy-dependent diffusion coefficient following $D(E_p) = \chi D_0 (E_p / E_0)^\delta$ for $E_p > E_0$, where $D_0 = 1 \times 10^{28}$ cm$^2$ s$^{-1}$ at $E_0 = 10$ GeV, $\delta$ depends on turbulence power spectrum, $t-t_i$ represents the diffusion time after escape. By integrating the contributions from all source points over the spherical shock surface, the explicit angular dependence on $\theta$ and $\phi$ is absorbed into a spherically symmetric form. The resulting local distribution of escaped protons at radius $R$ and time $t$ is written as \citep{2016ApJ...821...43Z}:
\begin{equation}
f(E_p,R,t)=\int_0^t
\frac{Q_0 E_p^{-p}}{4\pi^{3/2} R_{\rm dif} R_s(t_i) R}
\left\{
\exp\left[-\left(\frac{R-R_s(t_i)}{R_{\rm dif}}\right)^2\right]
-
\exp\left[-\left(\frac{R+R_s(t_i)}{R_{\rm dif}}\right)^2\right]
\right\}\,dt_i \, .
\label{eq:escape-local}
\end{equation}

To account for the finite extent of the illuminated gas, we further average the local proton distribution over a radial layer centered at $R_c$. We approximate the half-thickness of this illuminated layer by $R_{68}$, so that the volume-averaged proton spectrum at the SNR age $t_{\rm age}$ is
\begin{equation}
F_{\rm ave}(E_p,t_{\rm age})
=
{\int_{R_c-R_{68}}^{R_c+R_{68}} f(E_p,R,t_{\rm age}) \cdot R^2\,dR} \!\ /
{\int_{R_c-R_{68}}^{R_c+R_{68}} R^2\,dR}.
\label{eq:finite-volume}
\end{equation}

\begin{figure*}
    \centering
    \includegraphics[trim={0 0.cm 0 0}, clip,width=0.45\textwidth]{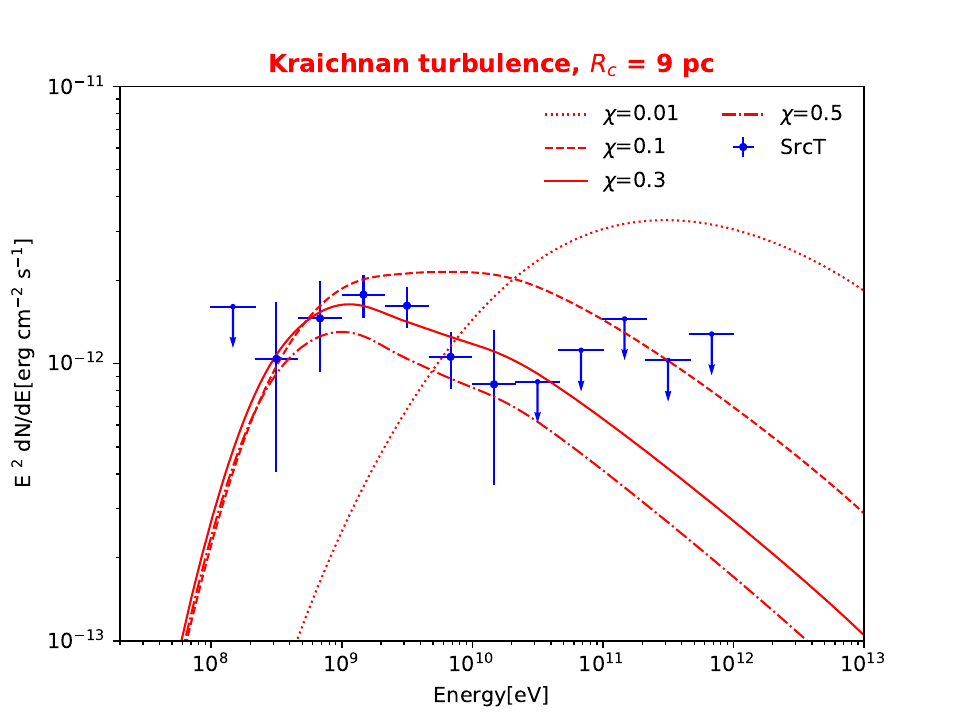} 
    \includegraphics[trim={0 0.cm 0 0}, clip,width=0.45\textwidth]{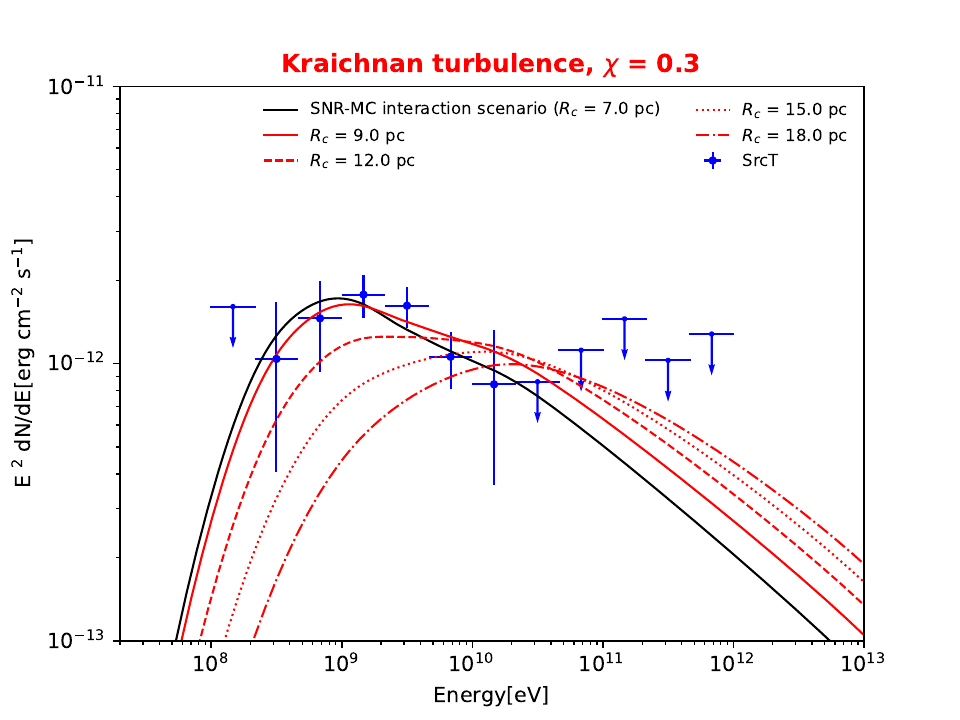} \\
    \includegraphics[trim={0 0.cm 0 0}, clip,width=0.45\textwidth]{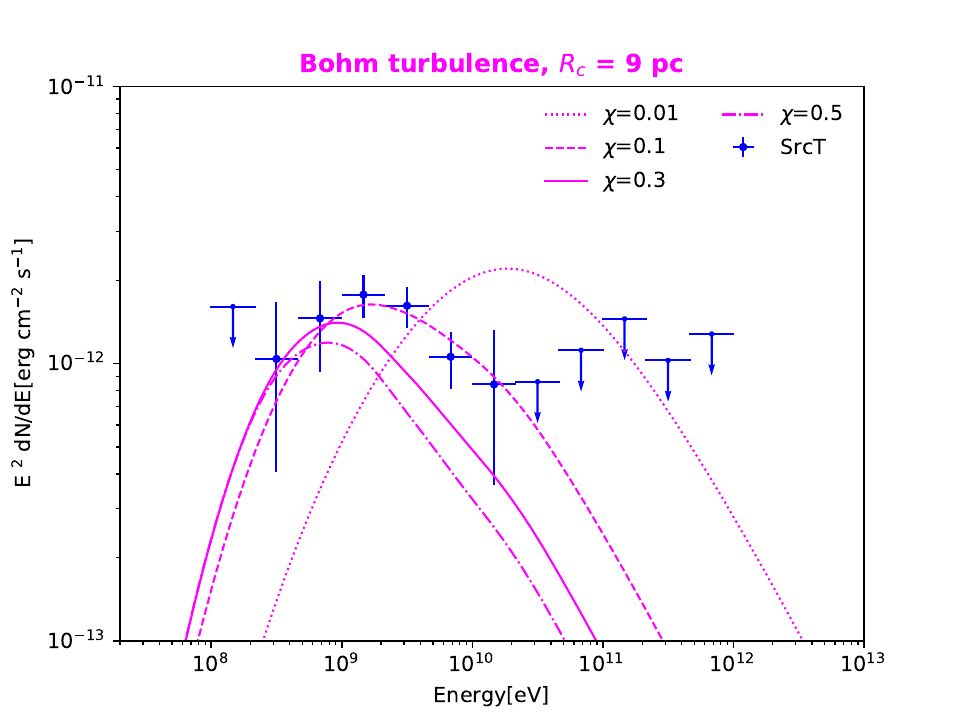} 
    \includegraphics[trim={0 0.cm 0 0}, clip,width=0.45\textwidth]{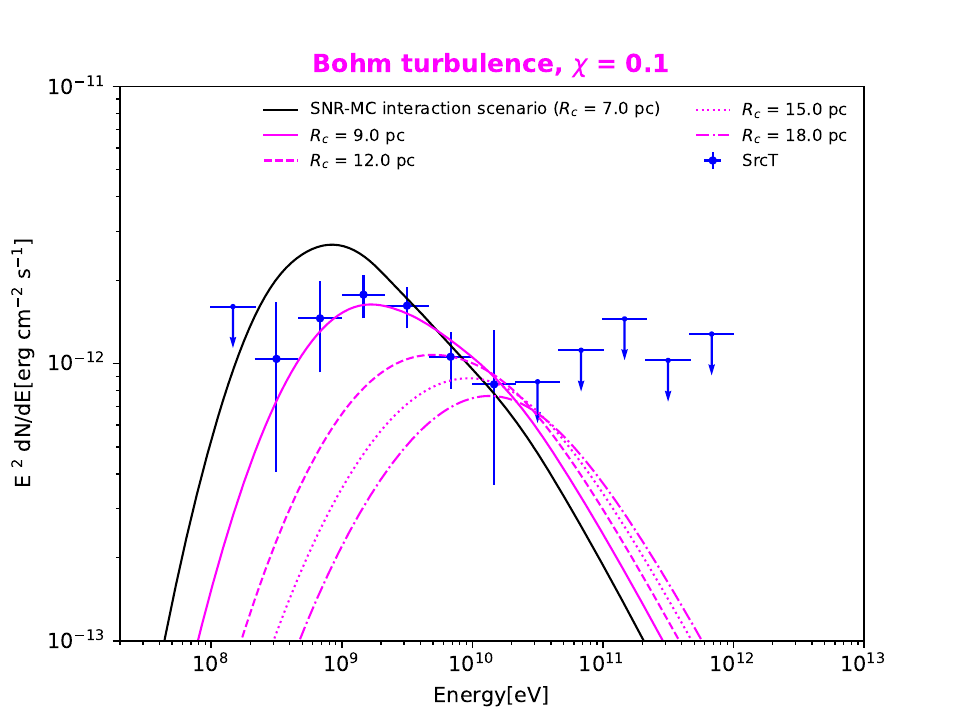} \\ 
    \caption{Hadronic modeling under parameters space. Top row: Kraichnan turbulence scenario, different fit line style corresponds to different $\chi$ and $R_{\rm c}$. Bottom row: Bohm turbulence scenario.}
   \label{fig:6}
 \end{figure*}

In this treatment, the radial integral is restricted to the illuminated layer, namely $R\in[R_c-R_{68},\,R_c+R_{68}]$, rather than the full extent of the parent molecular cloud. This choice is motivated by the assumption that only the near-side portion of the cloud is effectively illuminated by the escaping cosmic rays. The sketch of this model is shown in Figure \ref{fig:5}. Assuming that the remnant is in the Sedov phase, the observed shock radius of $R_{\rm s} \sim 7$ pc implies an age $\sim$ 3000 yr for a canonical explosion energy and ambient density \citep{1959sdmm.book.....S}. The injection spectrum is assumed to follow a power-law with an index of $p=2.0$. We consider two diffusion regimes characterized by different energy dependencies: $\delta = 0.5$ for Kraichnan turbulence \citep{1965PhFl....8.1385K} and $\delta = 1.0$ for Bohm diffusion \citep{1949bohm}, respectively (The Kolmogorov case ($\delta = 1/3$) is omitted because it yields much harder spectra indices). The distance between the SNR center and the molecular cloud is varied over $R_{\rm c} =$ 7, 9, 12, 15, and 18 pc, where $R_{\rm c} = 7$ pc corresponds to a direct interaction between the SNR shock and the cloud. The diffusion normalization factor is explored over $\chi = $ 0.01, 0.1, 0.3, and 0.5. The observed data are best reproduced in the Bohm-like case with $\chi \sim 0.1$. In contrast, the Kraichnan-like scenario can also provide acceptable fits, but typically requires a larger diffusion coefficient with $\chi \sim 0.3$ for certain parameter combinations. These results indicate that CR diffusion in the cavity surrounding the remnant is significantly suppressed compared to the canonical Galactic diffusion normalization $\rm D_0$ $\sim$ 3 $\times 10^{28}$ cm$^2$ s$^{-1}$ at 10 GeV \citep{2013A&ARv..21...70B}. The fit results are shown in Figure \ref{fig:6}. All fit lines require only a small fraction of the CRs injected by the SNR to reproduce the observed gamma-ray flux from the molecular cloud. For the estimated cloud density and distance, the proton energy required to account for the $\gamma$-ray emission from the molecular cloud is of order $W_p \sim 3\times10^{46}\ \rm erg$, which represents only a small fraction of the typical total CR energy budget of a young SNR ($\sim 10^{50}$ erg). This small CR energy is consistent with an escaped CR origin for the cloud emission, but by itself does not necessarily imply slow diffusion outside the SNR. Evidence for suppressed diffusion comes from the spatial and spectral modeling of the cloud emission.

The suppressed diffusion inferred here may reflect the magnetic and dynamical complexity of the cavity. Although the cavity is gas-poor, CR propagation is controlled mainly by magnetic-field geometry and turbulence. A spiral magnetic field inherited from the progenitor wind could increase the effective propagation path and lead to anisotropic transport. This possibility is motivated by the general expectation that rotating magnetized winds can generate wound-up magnetic-field structures \citep{1958ApJ...128..664P}. Previous massive-star activity or earlier supernova explosions may also have enhanced the turbulence level inside the bubble, thereby reducing the CR mean free path. Such collective effects of stellar winds, SNR shocks, and turbulence are expected to be important in superbubble environments \citep{2001SSRv...99..317B}. If the cloud lies within the near-source region affected by escaping CRs, the CR density gradient may remain higher than in the ambient ISM. In such a precursor-like environment, streaming CRs can excite Alfv\'enic turbulence through the resonant streaming instability \citep{1969ApJ...156..445K}, and scattering off these self-generated waves can reduce the CR mean free path and suppress the effective diffusion coefficient \citep{2016MNRAS.461.3552N}. Non-resonant CR-driven instabilities may also amplify magnetic fluctuations around SNR shocks \citep{2004MNRAS.353..550B,2013MNRAS.431..415B}. These effects provide plausible physical channels through which a low-density cavity can still exhibit inhibited CR transport.

\section{Conclusions}\label{sec:4}

The observed configuration of SNR G21.8-3.0, with a GeV $\gamma$-ray source spatially offset from the remnant and associated with a localized molecular cloud, supports a scenario in which CRs escaping from the SNR interact with external dense gas. The lack of significant $\gamma$-ray emission from the SNR shell, together with the presence of a surrounding low-density cavity, suggests that the escaped particles illuminate dense gas outside the remnant while most CRs remain confined near the source. The inferred diffusion coefficient, a factor of $\sim 30$ below the canonical Galactic value, points to a transport regime influenced by local magnetic-field structure and enhanced turbulence. 

In this context, the cavity should be viewed as part of the transport environment itself, rather than simply as an evacuated volume separating the accelerator from the target cloud. The suppressed diffusion may reflect the combined effects of local magnetic-field geometry, pre-existing turbulence, and CR-driven turbulence in the near-source region. This interpretation naturally explains the offset $\gamma$-ray emission and highlights the importance of circumstellar structures in shaping the early propagation of CRs. Our results suggest that an apparently evacuated wind-blown cavity should not be treated as a featureless free-propagation region, its magnetic and dynamical structures must be considered when interpreting the early transport of escaping CRs around SNRs.

\begin{acknowledgments}
This work is supported by the National Natural Science Foundation of China under the grants No. 12393852, 12393853, 12573047 and 12350610239. Yuan Li is supported by T.D.Lee Scholarship. We would like to thank Delia's invaluable discussions. This research made use of the data from the Milky Way Imaging Scroll Painting (MWISP) project, which is a multi-line survey in 12CO/13CO/C18O along the northern galactic plane with PMO-13.7m telescope. We are grateful to all the members of the MWISP working group, particularly the staff members at PMO-13.7m telescope, for their long-term support. MWISP was sponsored by National Key R$\&$D Program of China with grants 2023YFA1608000 $\&$ 2017YFA0402701 and by CAS Key Research Program of Frontier Sciences with grant QYZDJ-SSW-SLH047.
\end{acknowledgments}

\bibliographystyle{aasjournal}
\bibliography{ref}

\end{document}